\documentclass[conference]{IEEEtran}
\usepackage{cite}
\usepackage{amsmath,amssymb,amsfonts}
\usepackage{algorithmic}
\usepackage{graphicx}
\usepackage{textcomp}
\usepackage{xcolor}
\usepackage{tabularx,colortbl,caption}
\makeatletter
\newcommand{\linebreakand}{%
  \end{@IEEEauthorhalign}
  \hfill\mbox{}\par
  \mbox{}\hfill\begin{@IEEEauthorhalign}
}
\makeatother

\def\BibTeX{{\rm B\kern-.05em{\sc i\kern-.025em b}\kern-.08em
    T\kern-.1667em\lower.7ex\hbox{E}\kern-.125emX}}

\definecolor{iqcred}{cmyk}{0.07,1,0.67,0.31}
\definecolor{iqclightred}{rgb}{246,197,206}
\definecolor{iqcblue}{cmyk}{0.72,0.15,0,0.56}
\definecolor{iqclightblue}{rgb}{178,235,235}
\definecolor{iqcblack}{cmyk}{1,1,0,0.98}
\definecolor{Watblue}{cmyk}{0.9,0.48,0,0}	
\definecolor{Watred}{cmyk}{0.07,1,0.67,0.31}	
\definecolor{Watgreen}{cmyk}{0.72,0.15,0,0.56}
\definecolor{Watyellow}{cmyk}{0,0.09,0.80,0}
\definecolor{Sciblue}{cmyk}{0.9,0.47,0,0}
\definecolor{Mathpink}{cmyk}{0.04,0.9,0,0}
\definecolor{Engpurple}{cmyk}{0.52,0.7,0,0}
\definecolor{Watgray15}{cmyk}{0,0,0,0.15}
\definecolor{Watgray50}{cmyk}{0,0,0,0.5}

\newcommand{\corr}[1]{{\color{blue}{#1}}}

\begin{document}


\title{USEQIP: Outcomes and experiences from 17 years of undergraduate summer schools in experimental quantum information science}

\author{
\IEEEauthorblockN{John M. Donohue\textsuperscript{\textsection}}
\IEEEauthorblockA{\textit{Institute for Quantum Computing} \\
\textit{University of Waterloo}\\
Waterloo, ON, Canada \\
jdonohue@uwaterloo.ca}
\and
\IEEEauthorblockN{Michael J. Grabowecky\textsuperscript{\textsection}}
\IEEEauthorblockA{\textit{Institute for Quantum Computing} \\
\textit{University of Waterloo}\\
Waterloo, ON, Canada \\
mgrabowecky@uwaterloo.ca}
\and
\IEEEauthorblockN{George Nichols}
\IEEEauthorblockA{\textit{Institute for Quantum Computing} \\
\textit{University of Waterloo}\\
Waterloo, ON, Canada}
\linebreakand
\IEEEauthorblockN{Martin Laforest}
\IEEEauthorblockA{\textit{Quantacet Investments} \\
Sherbrooke, QC, Canada}
\and
\IEEEauthorblockN{Lino Eugene}
\IEEEauthorblockA{\textit{Quantum Nanofabrication and} \\
\textit{Characterization Facility (QNFCF)}\\
\textit{University of Waterloo}\\
Waterloo, ON, Canada}
\and
\IEEEauthorblockN{Fiona Thompson}
\IEEEauthorblockA{\textit{Institute for Quantum Computing} \\
\textit{and Dept. of English} \\
\textit{University of Waterloo}\\
Waterloo, ON, Canada}
\linebreakand
\IEEEauthorblockN{Peter Sprenger}
\IEEEauthorblockA{\textit{KUK Electronics} \\
Appenzell, AI, Switzerland}
\and
\IEEEauthorblockN{Kevin Resch}
\IEEEauthorblockA{\textit{Institute for Quantum Computing} \\
\textit{and Dept. of Physics and Astronomy} \\
\textit{University of Waterloo}\\
Waterloo, ON, Canada}
\and
\IEEEauthorblockN{David G. Cory}
\IEEEauthorblockA{\textit{Institute for Quantum Computing} \\
\textit{and Dept. of Chemistry} \\
\textit{University of Waterloo}\\
Waterloo, ON, Canada}
}

\maketitle

\begingroup\renewcommand\thefootnote{\textsection}
\footnotetext{Equal contribution}
\endgroup

\begin{abstract}

To grow the quantum information science and technology workforce, opportunities for students to gain experiential learning and build a sense of belonging in the broader community are essential. The Undergraduate School on Experimental Quantum Information Processing (USEQIP) is a two-week summer school for undergraduate students that has been held since 2009 with the goal of introducing undergraduate students from around the world to the tools of quantum information research, paired with a summer internship program. Here we report on the structure, impact, and outlook of the program, including hands-on laboratory activities refined over many iterations of the program. We highlight the career trajectories of program alumni, many of whom have made significant contributions to the quantum field.
\end{abstract}


\section{Introduction}

Quantum information science and technology (QIST) is a highly multi-disciplinary field of study that uses quantum phenomena such as superposition, interference, and entanglement to develop devices with enhanced selectivity, sensitivity, precision, efficiency, and security that ultimately exhibit superior performance compared with their classical counterparts. Individuals working on QIST research today span a wide variety of backgrounds including but not limited to physics, engineering, chemistry, computer science, mathematics, and nanoscience. QIST is a rapidly emerging discipline, with the Quantum Economic Development Consortium (QED-C) reporting that there are over 500 ``pure-play'' quantum companies worldwide with a workforce of over 14,000 quantum-exclusive workers, and a broader estimate of approximately 200,000 workers in roles relating to quantum in some capacity~\cite{QED-C}. This rapid growth has necessitated a significant shift in workforce demand from individuals highly specialized in a singular application of quantum technologies to individuals who are quantum-literate \textit{and} quantum-proficient: those who can advance quantum impact using a foundational set of tools and techniques. 

Many institutions have established fast-track higher education programs in quantum technologies to respond to this increased workforce demand. These programs range from summer schools to quantum information (QI) specializations in existing degrees, and there has been a sharp increase in dedicated programs entirely focused on quantum literacy worldwide \cite{aninder-Kaur, Aiello_2021}. Industry surveys on the state of the quantum workforce and training requirements for effective transition to the quantum workforce have widely reported the need for strong theoretical foundations with an equally strong emphasis on the necessity of hands-on experiences with quantum devices \cite{aninder-Kaur, Fox-Michael, Hughes, Greinert, Greinert-2024, QCanada}.  

To address these needs, we have developed and managed the Undergraduate School on Experimental Quantum Information Processing (USEQIP)~\cite{USEQIP}, a summer school program for undergraduate students that has run at the University of Waterloo through the Institute for Quantum Computing (IQC) annually since 2009\footnote{Excluding 2020 due to the COVID-19 pandemic, though the internship program associated with the program continued remotely. A limited 2021 program was adapted to run online\corr{;} all other iterations were fully in-person.}. In this article, we outline the structure and impact of this program in the landscape of the current quantum industry. In Sec.~\ref{sec:students}, we outline the student demographics and intended audience. We then overview the USEQIP curriculum in Sec.~\ref{sec:curric}, including lectures, labs, and social activities, as well as the associated research internship program. We then take a closer look at the hands-on lab activities and the infrastructure needed for them in Sec.~\ref{sec:labs}. Finally, we look at the impact of the program on students by examining their career pathways from post-workshop surveys in Sec.~\ref{sec:feedback}.

\section{Student Demographics}\label{sec:students}

From 2009--2025, USEQIP has hosted 349 in-person participants over 15 iterations (as well as 65 online during the COVID-19 lockdown in 2021). The initial class size of 11 students grew to a steady-state of 25-30 students per year by 2016. The limit of 30 students ensures that all individuals have sufficient hands-on time with equipment during the lab activities throughout the program.

USEQIP students are considered through an open application process, with selections based on academic performance, relevant experience, leadership activities, and enthusiasm for quantum information research, evaluated through a transcript, curriculum vitae, cover letter, and reference letter. If they apply for a research internship (see Sec.~\ref{sec:URA}), they are also evaluated based on research interest overlap with the research labs at IQC. Students are eligible if they are currently enrolled in an undergraduate program, with most successful students in their third year of studies or higher. From 2015--2025, an average of 250 applications were received per year (ranging from 198 to 294), with an average acceptance rate of 11\%.

Applications are considered from students both within and outside of Canada. Between 2019--2025, an average of 23\% of applications (406/1741) were domestic (from students at Canadian universities), and this group corresponds to an average of 47\% of invited program participants (94/205). Of the 414 total successful applicants from 2009--2025, students from Canadian universities were the most common (169), followed by the USA (126), India (37), China (11), the United Kingdom (10), and Germany (10). Overall, students from 35 countries have participated in USEQIP since 2009. The lack of a standard academic schedule internationally is a challenge for preparing an international summer school, and a likely reason for the relatively low attendance from European countries.

One of USEQIP's primary aims is to provide pathways to engage with QIST for students who may not otherwise have access. Students accepted to the program are not required to pay a registration fee and receive travel bursaries to cover flights, reducing financial barriers. We are also keenly aware of the under-representation of women and gender minorities in physics and other fields of study relevant to QIST~\cite{rangan2023status}. Since 2018, USEQIP has ensured that the first round of program acceptances are balanced between male and female among those who self-identify within the gender binary.

USEQIP is open to students from any field that has links to QIST, including physics, chemistry, engineering, computer science, and mathematics. While the summer program is heavily focused on experimental techniques in QIST, we strongly encourage participation from students without an experimental background or stated experimental research goals; their participation helps build a common language and understanding between theoretical and experimental concepts. Since 2020, applicants have been asked to self-identify their field of study between the options of ``Physics,'' ``Chemistry,'' ``Computer Science (CompSci),'' ``Mathematics,'' and ``Engineering,'' where they are allowed to pick more than one if appropriate. A majority of the applicants (62\%) and acceptees (68\%) identify as physicists, but a significant number identify as computer scientists (31\% of applicants, 19\% of acceptees), mathematicians (19\% of applicants, 19\% of acceptees), and engineers (19\% of applicants, 16\% of acceptees). Chemistry students are in a minority, with only 2\% of applicants and acceptees identifying with that field.

\section{Program Structure}\label{sec:curric}

\begin{table*}[htbp]
\begin{center}
\begin{tabularx}{\textwidth}{|>{\hsize=0.5\hsize}X|*{6}{>{\hsize=1.1\hsize\centering\arraybackslash}X|}} 

\hline
\rowcolor{lightgray}
\textbf{Week 1} & MON & TUE & WED & THU & FRI \\
 \hline
9-12 & \textbf{Lecture}: Quantum Mechanics Review & \textbf{Lecture}: Spin Qubits & \textbf{Lab}: Low-Temperature & \textbf{Lab}: NV Centres  & \textbf{Lecture}: Atomic QIP \\
\hline 
12-1 & Lunch Break & Lunch Break & Lunch Break & Lunch Break & Lunch Break \\
\hline
1-4 & \textbf{Lecture}: Quantum Information & \textbf{Lab}: NMR I & \textbf{Lecture}: Photonic Qubits & \textbf{Lecture}: Trapped-Ion QIP & Industry Guest Session \\
\hline
4-6 & Lab Tour I & Social I & Lab Tour II & Ball Hockey & Community Gathering \\
\hline \hline
\rowcolor{lightgray}
\textbf{Week 2} & MON & TUE & WED & THU & FRI \\
 \hline
9-12 & \textbf{Lecture}:  Quantum Algorithms &\textbf{Lecture}: Superconducting Qubits & \textbf{Lab}: Nanofabrication & \textbf{Lab}: QKD & \textbf{Lab}: NMR Challenge \\
\hline 
12-1 & Lunch Break & Lunch Break & Lunch Break & Lunch Break & Lunch Break \\
\hline
1-4 & \textbf{Lab}: Bell's Inequalities & \textbf{Lab}: Superconductors & \textbf{Lecture}: Quantum Error Correction & \textbf{Lab}: Ion Trap & \textbf{Lab}: NMR Challenge \\
\hline
4-6 &  &  & Applications Panel & \textbf{Lab}: NMR II & Closing Social \\
\hline 
 
\end{tabularx}
\end{center}
\caption{USEQIP schedule from 2023, presented as an example. Other years have slight variations but follow the same general trajectory. Note that students may experience the labs in another order, as they rotate through a sequence of three labs over the course of each week to stay within capacity limitations and maximize hands-on opportunities for the students.}
\label{tab1}

\end{table*}

The USEQIP program structure is intended to be accessible to undergraduate students with minimal experience in QIST. The majority of students have taken at least one course in linear algebra and quantum mechanics and have a significant interest in the field. Before starting the program, students are provided an introductory textbook on quantum computing~\cite{kaye2006introduction} and a set of practice problems. In recent years, to encourage students to work together on practice problems in advance and to facilitate students getting to know each other ahead of the program, we have prepared a Discord server open to students and instructors.

The exact structure of the program varies from year-to-year due to scheduling, equipment availability, and feedback-driven improvements to the program; a recent sample schedule from 2023 is included in Table~\ref{tab1}. Constant across iterations of the program are a sequence of lectures designed to support hands-on learning, special topics lectures designed to increase awareness of the breadth of QIST, and social activities designed to foster a sense of belonging among the students and with the broader quantum research community in Waterloo. The laboratory activities are detailed in Sec.~\ref{sec:labs}.

USEQIP is only possible because of the participation and willingness of the student, staff, and faculty communities at IQC. The USEQIP program today is organized by a program manager and a lab manager, both with a background in QIST, and accounts for 8-10\% of their full-time workload. To build, maintain, and facilitate the labs, two other technical staff members are heavily involved. An events coordinator and visitor program manager are also heavily involved to assist with immigration, student residence, catering, social activities, hiring (for the internship program), and other logistics, adding up to approximately 100~hours of administrative and organizational work. Graduate students are recruited to assist with the lab sessions, ensuring a student-to-facilitator ratio of 5:1 at most. A strong institutional culture of supporting community outreach has been essential for recruiting enough volunteers to run these activities, with a small stipend provided in appreciation of their time. We attempt to have a mixture of junior and senior graduate students when possible to help keep the activities consistent from year-to-year as students graduate. Most lectures are designed and taught by either the program/lab manager, faculty, or senior post-doctoral fellows, with most IQC faculty having contributed labs or lectures at some point through the program's history. All slides are provided to students for review and archived as resources for future iterations of the summer school.

\subsection{Lectures}

Lectures during USEQIP are structured to support the lab activities that follow or to build upon ideas introduced in prior labs. Most are delivered by faculty members and lecturers at the University of Waterloo, though occasionally will be delivered by senior graduate students and post-doctoral fellows. A standard lecture session is three hours long with a 15-minute break halfway through, with interactive elements and problem solving strongly encouraged. The pedagogical structure follows a spins-first approach to quantum mechanics~\cite{sadaghiani2016spin,knill2002,Ramanathan2004} before generalizing to other implementations.

USEQIP opens with a review of the basic linear algebra and quantum mechanics techniques necessary for understanding most QIST fundamentals. This includes bra-ket notation, pure states, projective measurements, observables, unitary operations, the tensor product, entanglement, and density matrices, with a brief introduction to decoherence. While this session is review for many students, it is essential for establishing a common language and notation for students who come from different academic backgrounds. The toolbox developed in this session is applied in a follow-up lecture on quantum information, introducing the circuit model and principles such as state discrimination, superdense coding, quantum teleportation, and the no-cloning theorem.

To connect these concepts to a tangible system, we introduce spin as a first physical qubit in the context of liquid-state nuclear magnetic resonance (NMR). Despite scalability limitations, liquid-state NMR provides a reliable, available, and concrete experimental platform for exploring quantum control. After connecting spin rotations to previously-introduced unitary transformations, we guide students through the Bloch equations for spin evolution, control via RF pulses, the rotating frame of reference, and how to interpret the time evolution of the spin state through the free-induction decay (FID) signal. In addition, this lecture introduces two-qubit systems and dynamics, including the definition of the scalar coupling interaction, a description of correlated and uncorrelated states in two spins, and the difference between observable and unobservable states in quantum mechanics. We then introduce techniques such as coherence transfer and pseudo-pure state preparation~\cite{PPS} and provide a brief introduction to nitrogen vacancy (NV) centers in diamond systems as another spin-based pathway to a qubit. In general, this lecture directly precedes the first lab activities, where NMR experiments are performed to directly observe one- and two-qubit dynamics.

We present photonic qubits as the second tangible system, where we connect the Bloch sphere picture introduced for spin qubits to photon polarization and introduce tools like wave-plates,  beam-splitters, polarizers, detectors, and coincidence logic for manipulating and measuring photons. We then focus on polarization-entangled photon sources, introducing spontaneous parametric down-conversion and geometries such as the two-crystal ``sandwich'' source~\cite{kwiat99spdc} and the Sagnac source~\cite{kim2006phase}, as well as effects like Hong-Ou-Mandel interference~\cite{HOM} and Bell state measurement. Next, we introduce Bell's inequalities and the CHSH test of non-locality, which students will later investigate in the laboratory.

Other experimental platforms for QIST are introduced afterwards. We introduce superconducting qubit circuit architectures through quantization of the LC-oscillator and the Cooper-pair box, following a discussion on the properties of Josephson junctions and a laboratory on low-temperature physics. Trapped-ion systems are introduced via the Paul trap and optical qubit control. We include a lecture on quantum algorithms, introducing computational efficiency and archetypal examples such as Deutsch's, Simon's, Shor's, and Grover's algorithms. We also introduce students to quantum error correction, detailing the nine-qubit Shor code and introducing stabilizers and the Toric surface code. A lecture on either the theory of quantum cryptography or experimental realizations of long-distance quantum communication channels is often included as well.

Depending on the availability of lecturers, many special topics lectures are included to provide a broader view of quantum information science. One-off special topics in the past have included quantum optomechanics, light-matter interactions, neutron optics, quantum simulation, open quantum systems, quantum dots, relativistic quantum information, and more. We have also included panel discussions on applications or career opportunities to increase the breadth of voices present and to connect to the quantum industry.

\subsection{Social activities}

A significant amount of time is dedicated to social activities to support the program's goals of community-building and increasing feelings of belonging in QIST research. Specific activities vary from year-to-year, but have included bowling, ball hockey, restaurant visits, and amusement parks. For the weekend in the middle of USEQIP, we organize an excursion to Niagara Falls, a notable nearby landmark. The USEQIP students are also invited to an institute-wide community gathering during the program to meet graduate students and faculty members, providing an opportunity for networking.

\subsection{Student research internships}\label{sec:URA}

In conjunction with the summer school, applicants are encouraged to apply for undergraduate research awards (URAs) and stay on campus for the full summer for paid research experience in an academic research group. Over the course of the program, 267 research internships have been organized. From 2022-2025, over 91\% of USEQIP students (102/112) secured an internship. The relationship between the USEQIP and URA programs is symbiotic: The USEQIP summer school ensures that these interns have opportunities to develop social ties with their peers early in their stay and provides a crash-course in many sub-fields of QIST, and the URA program alleviates concerns potential USEQIP students may have about taking time off for other summer work.

\section{Laboratory Activities}\label{sec:labs}


\subsection{\label{NMR} Nuclear magnetic resonance (NMR)}


\begin{figure}
    \centering
    \includegraphics[width=\columnwidth]{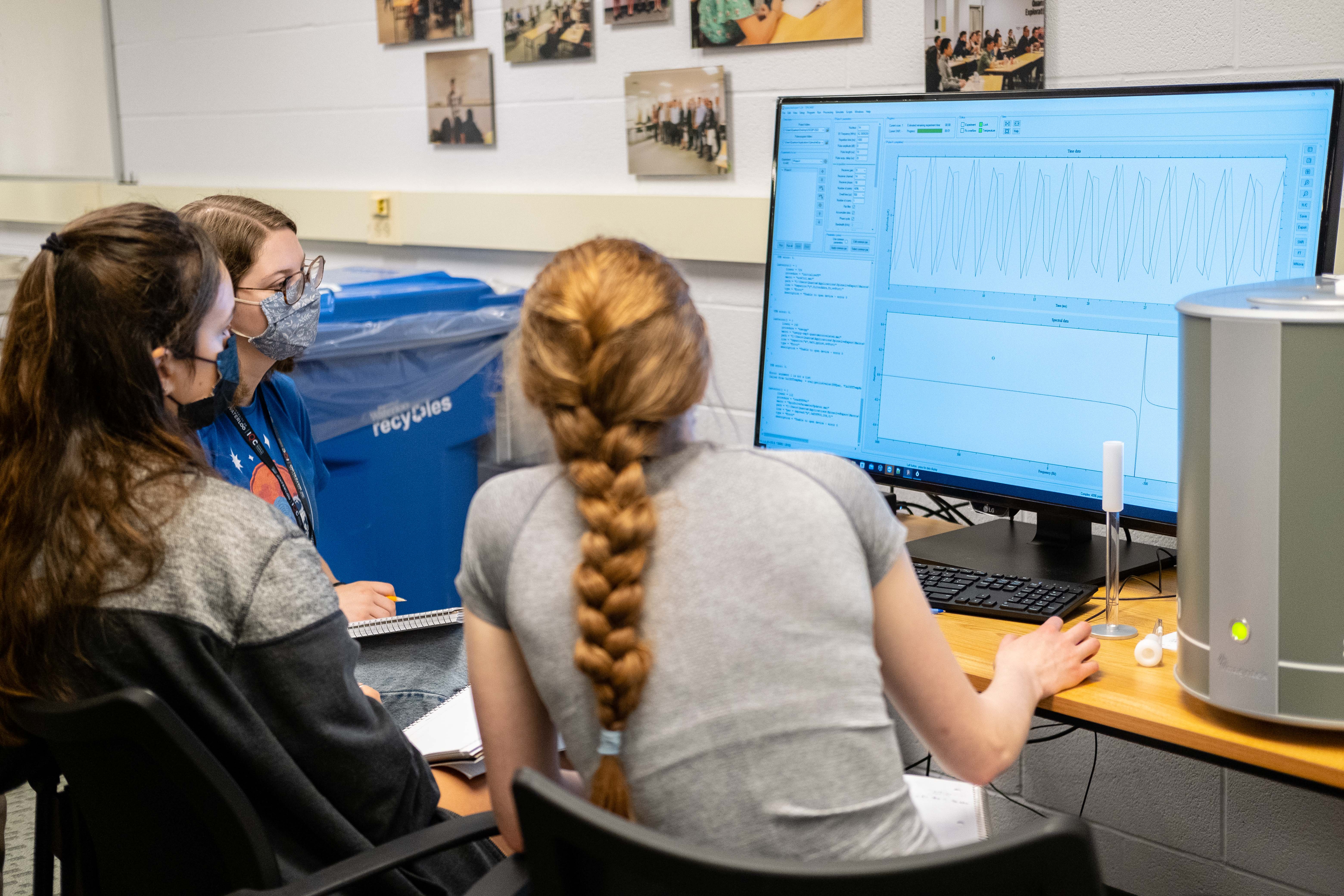}
    \caption{Students observe a free induction decay (FID) trace on a desktop NMR spectrometer during USEQIP 2022.}
    \label{fig:labphotos-NMR}
\end{figure}

Desktop nuclear magnetic resonance spectrometers can be implemented as small-scale quantum processors, allowing students to explore coherent control techniques~\cite{hou2021spinq, PRICE1999371}. These NMR systems are very robust, easy to maintain, relatively affordable, and operate at room temperature, making them an accessible platform for the exploration of quantum information processing (QIP). In USEQIP, students are guided through a set of experiments in which they learn how to characterize and control one- and two-qubit systems. They use two different samples: water (H$_2$O, one-qubit), and 13C-enriched chloroform ($^{13}$CHCl$_3$, two-qubit). Students learn how to implement and calibrate basic quantum gates, create correlations between multiple qubits, and gain an intuition for noise processes limiting coherence times in quantum systems. The supporting documentation and the lab instruction is delivered in the language of QIP rather than the language of NMR to better emphasize the isomorphism between various modalities of quantum computing. Experimental observations are related back to two-level quantum system dynamics (Bloch equations), providing an easy way for students to compare with theoretical expectations. The focus on control techniques provides students with a working understanding of how to deploy the NMR spectrometers as a two-qubit quantum processor and general knowledge that can be applied to other QIST implementations.

Key experiments include: Ramsey spectroscopy for identifying qubit resonance frequencies and coherence times ($T_2^*$); Rabi oscillations for calibrating control parameters; measurement of the coupling strength between two qubits; utilizing coupling to obtain correlated states between two qubits; an exploration of observable versus unobservable quantum states; and methodology for how to simulate pure state dynamics at room temperature (pseudo-pure state preparation).

Students are given access to an NMR simulator built in Wolfram Mathematica, which is capable of demonstrating the same experiments as those conducted in the lab. This simulator facilitates further exploration after the nominal lab time and is a useful tool the students can make use of during the NMR challenge day, in which they choose a topic in QIP and attempt to implement it on the NMR spectrometers (discussed further in Sec.~\ref{NMRchal}). The set of experiments described here serves as preliminary knowledge necessary to enable student exploration on the NMR challenge day.

Early iterations of USEQIP had students use conventional high-field NMR spectrometers. This approach gave students experience with manual techniques for experimental setup, such as shimming to improve the field homogeneity across the sample, but often meant that little time was left for the actual quantum control elements. By shifting to desktop NMR spectrometers, students are able to spend more time understanding the essential elements of quantum control without getting tied up in the details and language of NMR spectroscopy. The comparatively lower cost and maintenance demands of desktop devices also allows more devices to run in parallel, increasing the hands-on time students have with the equipment.

\subsection{Quantum key distribution (QKD)}

\begin{figure}
    \centering
    \includegraphics[width=\columnwidth]{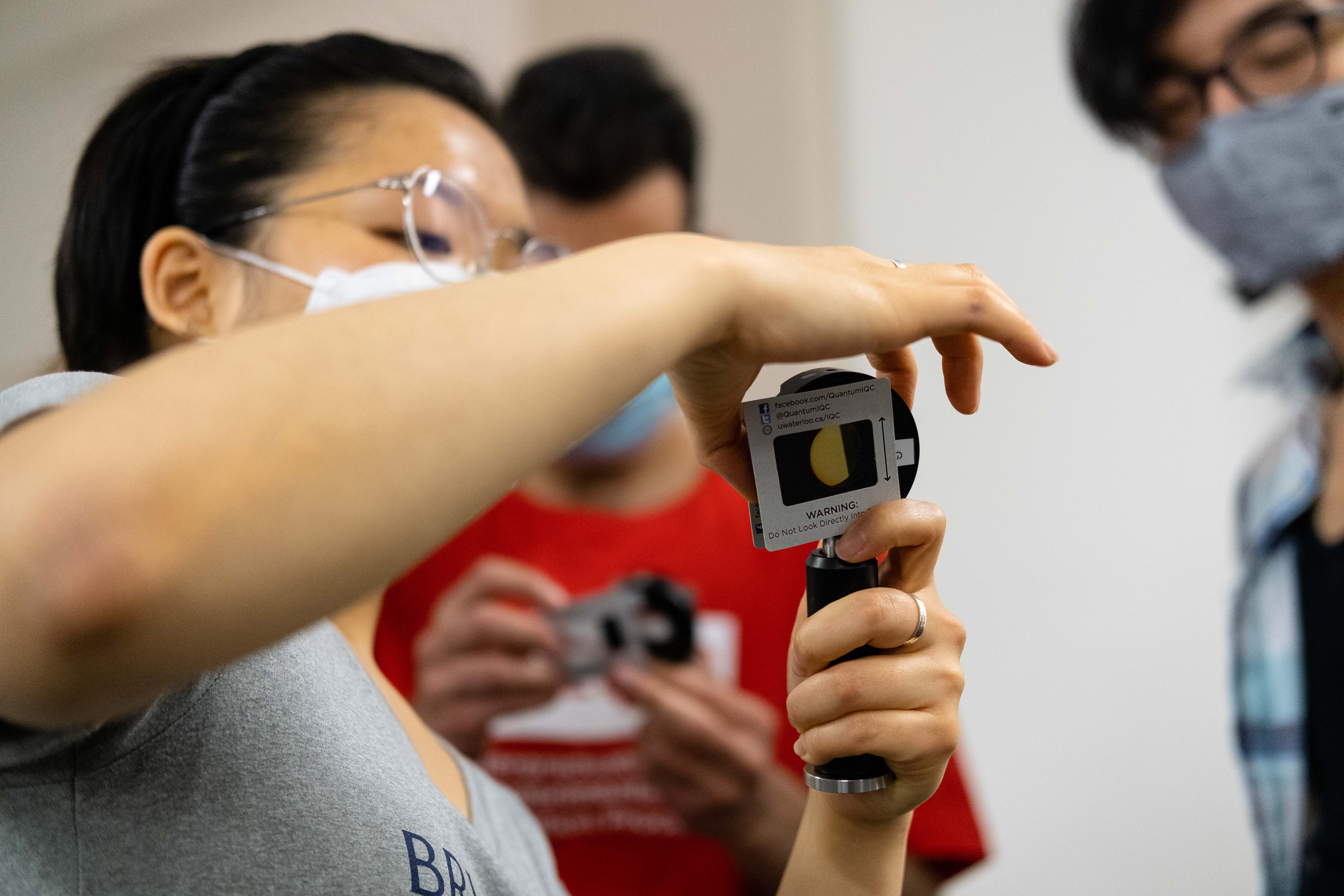}
    \caption{Students use polarizing slides to identify unknown optical components as part of the QKD lab during USEQIP 2022.}
    \label{fig:labphotos-QKD}
\end{figure}

Quantum key distribution (QKD) is introduced to students through an analogy kit, a form often used to introduce students to key ideas in quantum information science such as randomness, superposition, and measurement bases, including at the high-school level~\cite{utama2020hands}. The QKD analogy kit uses a unattenuated laser diode in place of a quantum light source, rendering it ineffective at genuine secure key exchange as the signal may be easily copied or split. However, by encoding the analogous polarization states into the laser beam and analyzing the output signals from the detection photo-diodes, it is possible to mimic the behaviour students would see from a genuine single-photon source.

The USEQIP QKD lab activity is presented as an engineering challenge rather than a pure key generation challenge to fill a three-hour laboratory session. Students are given a brief introduction to QKD and tasked with building a QKD analogy kit using standard optomechanical components. However, the USEQIP students are not given immediate instruction on what each component is, and the components are intentionally unlabeled. Students are given two polarizing slides as trusted resources and must determine the function of the other components using those, as seen in Fig.~\ref{fig:labphotos-QKD}. This includes distinguishing half-wave plates from other polarization rotators, determining the functionality of polarizing beam-splitters compared to plate polarizers in an optical setup, identifying the optical axes of wave plates, and strategically developing an optimal layout with limited breadboard space. Facilitators are instructed to give only limited hints to the students during the first half of the session, with the exception of how to connect the light sensors to an Arduino and GUI for data analysis.

This approach forces students to think critically about the needs of the experiment. Some students with experience in optics labs have an advantage with optomechanical components, but most will be using these relevant tools for the first time. After constructing the experiment and verifying the behaviour is as expected, students use their setup to distribute a key over an eavesdropper-free channel by role-playing as Alice and Bob. They then must distinguish between a series of further unlabeled components which potentially act as an eavesdropper, where they are challenged to determine if the component performs an eavesdropping attack and which basis it measures in.

\subsection{Entangled photon pairs and Bell's inequality}

USEQIP students use a source of polarization-entangled photons to explore entanglement, Bell's inequalities, remote state preparation, and quantum state tomography. The purpose of this lab is to deepen the student's understanding of entanglement and non-locality, as well as connect experimentally accessible measurements, such as photon counts, to models like expectation values. By the end of the laboratory, the student should have an understanding of how to account for statistical variations in quantum measurements.

Students are given access to a Sagnac-style source of polarization-entangled photon pairs pumped by a 405~nm continuous-wave laser to produce nearly-degenerate photon pairs at 810~nm. The system is prepared before the students arrive to produce one of the four canonical Bell states with a reasonably high fidelity (95\% or greater)~\cite{kim2006phase}. The source is enclosed, but the output photon pairs are fibre-coupled to student-accessible measurement setups consisting of a quarter-wave plate, half-wave plate, and polarizing beam-splitter, with fibre-coupled outputs to an avalanche photo-diode detector and coincidence logic. After a brief introduction connecting photon detections to projective measurements and defining how to calculate the error in measurements based on the Poissonian noise inherent to photon counting, students must determine which Bell state is being produced by the source and make the 16 projective measurements required to test the CHSH inequality~\cite{CHSH}. They use those measurements to calculate the Bell parameter $S$ and its uncertainty, forcing them to consider not just the violation of the inequality but the statistical certainty as well. They are challenged to manipulate variables to obtain the most statistically significant violation of the CHSH inequality compared to the other groups by adjusting the lighting in the room, the time that photon counts are accumulated for, and the intensity of the laser.

Next, students explore quantum state reconstruction using linear inversion tomography~\cite{james2001measurement}. First, they use Alice's measurement apparatus to remotely prepare Bob's qubit in a particular one-qubit state~\cite{bennett2001remote} and collect four measurements to reconstruct the density matrix of Bob's remotely prepared qubit. By performing tomography using both the detections from Bob's detector apparatus alone and the coincidence between Alice's and Bob's detectors, students can see the effect of remote state preparation\corr{:} the state measured via coincidence is highly coherent, while Bob's state independent of Alice is nearly maximally mixed. If time allows, students perform the 16 measurements needed for two-qubit tomography and perform linear inversion for the 4-by-4 two-qubit density matrix. Students are usually able to perform the one-qubit linear inversion by hand with some guidance, while the two-qubit inversion benefits from computational tools or mathematical software to complete in the three-hour lab window.

Some versions of this activity throughout USEQIP's history have incorporated other optics experiments, such a polarization-based implementation of the quantum Zeno effect.

\subsection{Nitrogen-vacancy (NV) centres}
The electrons trapped within the nitrogen-vacancy (NV) center defect in a diamond crystal lattice are widely deployed as a hybrid spin-optical QIP system with applications in quantum sensing due to their high sensitivity to applied magnetic fields. Although the magnetic field sensitivity for NV centers is lower than superconducting quantum interference devices (SQUIDs), NV centers offer superior spatial resolution and do not require cryogenic temperatures to operate~\cite{sensitivity}. Instead, laser radiation is used to drive electron transitions through the inter-system crossing, a phonon mediated process enforced by quantum selection rules, to prepare the electronic spins in a near-pure quantum state at room temperature. The fluorescent properties of the vacancy electrons enable optical readout of the quantum state. Coherent control can be implemented by pulsing resonant radio frequency (RF) radiation. Low-cost setups for continuous wave optically detected magnetic resonance (ODMR) with NV centres have been deployed for introducing students to quantum sensing techniques~\cite{haverkamp2025low}. Our setups include pulsing capability for the microwave and laser excitation, requiring higher grade RF electronics and phase stable RF sources compared with these lower cost alternatives but enabling a a wider range of experiments.

In this USEQIP lab activity, students demonstrate coherent control with an ensemble NV system by optically preparing a quantum state, manipulating the spin state using pulsed microwave radiation, and optically reading out the quantum system through a measurement of relative fluorescence under laser excitation. They utilize this control to perform basic quantum gates (creating superpositions and performing bit flips) and identify the optimal measurement time to read out the state of the NV ensemble qubit. In contrast to the NMR experiments, in which the inner workings of the systems are largely hidden away from the students, the NV apparatus provides the students with an opportunity to see and manipulate every component of a spin system. Once control is demonstrated, students deploy the NV ensemble as a magnetic field quantum sensor. Here, the students use continuous wave (CW) spectroscopy to measure the magnetic interactions of the vacancy centers with surrounding nitrogen-14 and carbon-13 spins and explore the effects of applying direct current (DC) magnetic fields at various orientations relative to the principle axis systems of the NV ensemble. The strengths of the applied magnetic fields are directly measured from the strength of the Zeeman interaction observed in the CW spectroscopy experiment. 

\subsection{Low-temperature physics and superconductivity}


Superconducting qubits are a leading quantum information implementation with widespread academic and commercial development. Such systems rely on cryogenic temperatures to reach the superconducting transition states of their materials (typically 1K-10K) and to isolate them from thermal noise ($<100$ mK for negligible thermal blackbody photon population at few-GHz frequencies).
 
This lab first introduces basic concepts of transport measurement, superconductivity, and building blocks of superconducting quantum circuits, as well as general low-temperature experimental skills. Students build a thermometer by soldering a carbon resistor with twisted-pair wires in a four-point measurement configuration. This resistor is then calibrated by measuring at room temperature and immersed in liquid nitrogen and fit to a simple exponential function before students are given a ‘mystery solution’ of dry ice in acetone and asked to measure its temperature.

Concepts of superconductivity introduced in lectures are then supported by an exploration of the Meissner effect, demonstrated by flux pinning and levitation of small neodymium magnets above a pucks of YBCO superconductor cooled with liquid nitrogen. Students dispense their own liquid nitrogen after a safety briefing and perform the demonstration independently. The perfect diamagnetism of the superconductor is connected to physics concepts students are already familiar with via a brief demonstration of Lens’ law in poor and good conductors. Magnetic field viewing film can be used to image pinned flux in the YBCO pucks by pressing a magnet onto its surface and confirming that this pinned field disappears on warming.

Partway through the session, the students transition to measuring the transport properties of a superconductor-insulator-superconductor (SIS) Josephson junction (niobium, cooled to a base temperature of 1.5K using variable-temperature liquid helium flow cryostats), which are the beating hearts of any superconducting qubit architecture, as they provide the necessary non-linear inductance creating the anharmonicity required to define a two-level system. Students explore the AC and DC current-voltage characteristics of the junctions, cementing the importance of 4-point measurement along the way. They begin by characterizing the DC properties of Josephson junctions, experimentally demonstrating the DC Josephson effect by quantitatively measuring the critical current, gap voltage, and normal state resistance of the junctions. The lab session ends with a demonstration of the AC Josephson effect in which the junctions are driven with RF radiation, producing quantized (Shapiro) steps in measured voltage (depending only on fundamental constants $h$ and $e$, as well as the applied RF frequency). These quantized voltage steps are employed in quantum metrology and are the quantum voltage standard developed by NIST~\cite{NIST}. 

In previous years, demonstrations have included superfluidity as a directly observable condensate phase. Liquid helium is transferred into a glass dewar and pumped to cool below the lambda transition. The superfluid transition is apparent when vigorous boiling of the helium bath stops abruptly due to the extreme thermal conductivity of the superfluid. The zero-viscosity property is demonstrated by lowering a glass tube plugged with a nanoporous ceramic filter first into the normal fluid, demonstrating that the tube holds liquid, then into the superfluid where a rapid flow through the ceramic is observed.
Superconducting Low-inductance Undulatory Galvanometers (SLUGs) \cite{ Clarke01011966} have also been used to demonstrate the DC and AC Josephson effect with students assembling their own SLUG devices. This functions identically to the demonstration with nanofabricated devices with the advantage of being more ‘hands-on’, but has reliability issues.
Superconducting resonators have been used to introduce frequency domain measurements with vector network analyzers. The frequency spectrum of a superconducting resonator is measured as a function of temperature to determine its temperature-dependent resonant frequency and quality factor. This can then be related to the two-fluid model of condensates, and concepts of nonlinearity and kinetic inductance are discussed.

\subsection{Nanofabrication}\label{sec:labs:nanofab}

The USEQIP students are introduced to micro- and nanofabrication with a one-hour lecture followed by a three-hour fabrication session in the main cleanroom of the Quantum Nano-Fabrication and Characterization Facility (QNFCF). In the cleanroom, they are shown photolithography, plasma etching of aluminum, and scanning electron microscopy as part of the fabrication of quantum devices based on Josephson junctions. Students are required to take cleanroom safety training before the session. After gowning up, we divide the students into groups of 3-4 students to explore three stations with tools applicable to many fabrication processes in quantum technologies.

\begin{figure}
    \centering
    \includegraphics[width=\columnwidth]{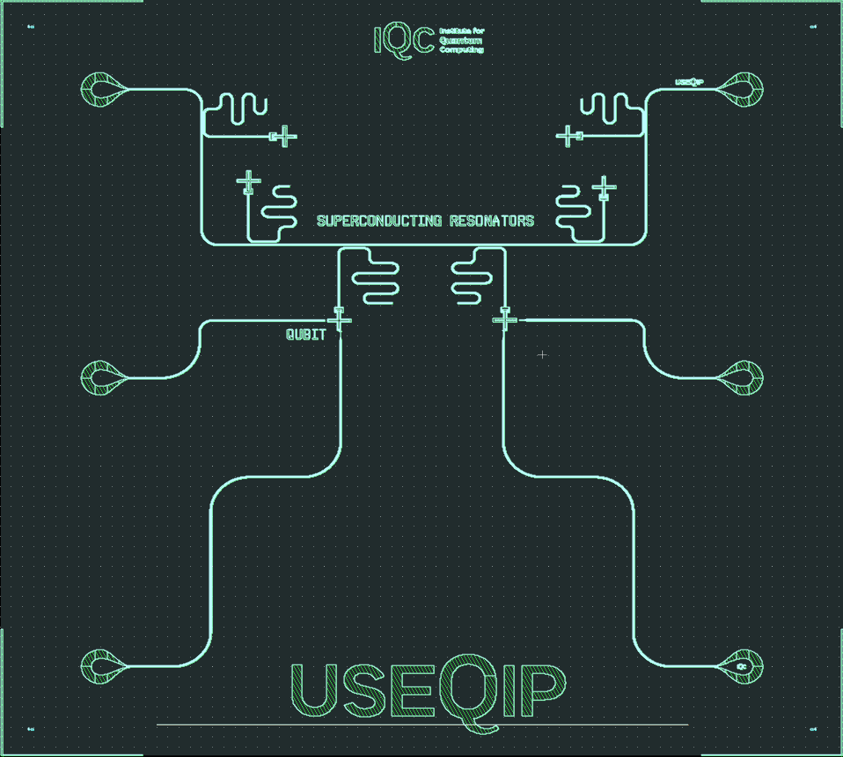}
    \caption{ Design of the chip developed for the USEQIP 2022 nanofabrication lab activity.}
    \label{fig:labphotos-QNFCF}
\end{figure}

Prior to the students' arrival, prime Si wafers are coated with 100~nm Al deposited in an Angstrom NexDep e-beam evaporator.  In one module, the students are shown and explained the steps of photolithography. A Shipley S1811 photoresist is spun onto the wafer in a CEE 200X spinner, baked in a CEE 1300X hotplate then exposed with an Heidelberg MLA 150 maskless aligner. The chip layout used for the exposure, seen in Fig.~\ref{fig:labphotos-QNFCF}, consists of two transmon qubits with coupled microwave resonators similar to devices used in superconducting quantum information processing. After exposure, participants are given the opportunity to develop the photoresist by immersing the wafer in MF-319 developer followed by a rinse in deionized water. The exposed patterns are inspected under an optical microscope. In another module, the students are quickly introduced to physics of plasma and the importance of plasma etching in nanofabrication, especially for quantum devices. The photoresist patterns are transferred into the aluminum layer by plasma etching with a gas mixture of Cl2/BCl3 in an Oxford Plasmalab 100 system equipped with an inductively coupled plasma (ICP) tube of 380 mm. In the final module, the etched structures are inspected under a JEOL JSM-7200F scanning electron microscope (SEM) where each student tries to get the best image possible by adjusting the focus. As a surprise, the names of the students can be found inscribed on the chip under the SEM.


After the lab activity, QNFCF staff dice the completed wafers so that students can take home a chip as a souvenir. The experience of the cleanroom environment in the QNFCF is often noted as a highlight by students, both for the access to specialized equipment and the unique environment it takes place in.


\subsection{\label{NMRchal} NMR challenge}
In the last day of the program, students are split into groups of three to five and are tasked with researching and implementing a topic in QIP using the NMR spectrometers. The morning is focused on brainstorming ideas, researching implementations, and discussing feasibility with instructors. The afternoon provides the students with an opportunity to test their implementations, with time allotted for a brief  presentation of each group's findings at the end. The preliminary NMR experiments outlined in Sec.~\ref{NMR} provide students with the baseline knowledge of coherent control, allowing the focus to shift to the applications of QIP.

Some topics explored by students include quantum simulation (truncated quantum harmonic oscillator with a two-spin system), Spinor dynamics, Bell's inequality violation, the Deutsch-Josza algorithm, Grover's algorithm, noise characterization, composite pulse engineering, randomized benchmarking, implementations of specific quantum gates (examples include CNOT and Hadamard gates), and more.

The event highlights the versatility of NMR for exploring QIP techniques. There is a focus on topics that are portable between different physical platforms of quantum computing; we avoid topics that are specific to NMR. It is important to note that it is often the case that a group of students fails to successfully implement their topic on the physical devices, which teaches them an important aspect of the scientific process. Although it is important to understand how to implement an experiment correctly, it is just as important for students to understand where an experiment might have gone wrong, and for them to draw conclusions on how to improve an implementation given the opportunity in the future. 

\subsection{Other labs and activities}\label{sec:labs:other}

Over the course of USEQIP's history, many other labs have been included in the program. For example, in recent iterations, a linear Paul trap designed for levitating dust spores~\cite{mcginness2019submitter,saito2024measurement} has been included for students to explore technologies related to trapped-ion quantum computing. Students control the AC and DC voltage in the trap to examine the trapping effectiveness and use a Jupyter notebook to calculate the necessary voltages and AC frequencies for single-ion experiments. Other labs have explored lock-in amplifiers, quantum-dot photon sources~\cite{pennacchietti2024oscillating},  atomic force microscopy, microwave electronics, scanning tunneling microscopy, and hacking QKD systems. Guest sessions on specific industry platforms have been delivered from IBM Quantum, Anyon Systems Inc., and Xanadu.

To highlight areas of experimental research not formally covered in the summer school, we hold lab tours throughout the program, showing students both specific research labs and the experimental infrastructure required for quantum research.

\subsection{Further uses of lab infrastructure}

In the first many iterations of USEQIP, lab activities took place in spaces otherwise used for research. As the program has evolved, we have been able to develop dedicated infrastructure strictly for educational purposes, including NMR spectrometers, NV apparatuses, entangled photon sources, liquid helium flow cryostats, and more, housed in an educational environment dubbed the Quantum Exploration Space. This reconfigurable space allows for lectures and labs to be held in the same location and for lab activities to be set up and move as needed to run in parallel.

Having this infrastructure available for educational purposes has allowed us to use them for programs outside of USEQIP, informed by lessons learned from many years of the USEQIP program. A one-year Masters' program in Physics with an emphasis on quantum technology was developed with three lab-focused courses using some of the tools developed initially for USEQIP. These lab activities have also been integrated into undergraduate physics lab courses and quantum engineering courses at the University of Waterloo and used in one-off events such as undergraduate quantum clubs, and hands-on workshops in partnership with local start-up incubators.

Simplified versions of the experiments have also been run for earlier learners. The NMR spectrometers and entangled photon sources have been integrated into programs for high-school students, notably the Quantum School for Young Students (QSYS), where students use these tools to connect abstract circuit models to tangible physical systems~\cite{laforest2017bringing}. This infrastructure has allowed us to perform live demonstrations of quantum effects like spin resonance in NV and NMR experiments and Bell's inequality violations for groups of high-school teachers as part of the Quantum for Educators (QEd) program~\cite{laforest2017bringing,donohue2026outreach}.

\section{Participant Feedback and Impact}\label{sec:feedback}

\subsection{Post-workshop surveys}

We evaluate participant feedback and program impact through detailed post-workshop surveys, occasional alumni surveys, and alumni tracking. We use this feedback year-over-year to improve labs and overall flow of the program, including logistical needs such as immigration support, catering, and accommodations. For USEQIP 2013--2025, surveys on the previous week's material were sent at the end of each week to ensure the students' memories are fresh. Response rates to post-surveys are generally quite high, ranging from 52\% (13/25 in 2025) to 74\% (23/31 in 2018). 

To evaluate the labs, attendees were asked questions about each lab on a 4-point Likert scale, ranging from Strongly Agree (4 points) to Strongly Disagree (1 point). From 2017--2025, for each lab activity, students were asked for their agreement with phrases regarding interest (``I found the lab interesting'') and relevance (``The lab helped clarify lecture content''). From 2023--2025, a question was added about novelty (``The lab was unlike activities I have seen before''). The average ratings are presented in Table~\ref{tab:labfeedback} and show that a strong majority agree or strongly agree with these statements about all labs. Students were provided space for open-ended feedback and provided responses such as \textit{``The labs were an amazing exposure to actual experiments that are conducting [sic] in current research. It really helped strengthen the theory with actual experimental observations.''} (Anonymous 2023).

\begin{table}
    \centering
    \begin{tabular}{c|c|c|c|c}
         & Interest  & Relevance  & Novelty & N \\
         & (/4) & (/4) & (/4) & \\
        \hline
        NMR & 3.48 & 3.65 & 3.78 & 118 (44)\\
        Low-temperature & 3.68 & 3.38 & 3.58 & 118 (44)\\
        NV & 3.28 & 3.25 & 3.67 & 101 (44)\\
        QKD & 3.71 & 3.61 & 3.42 & 118 (44)\\
        Entanglement & 3.47 & 3.48 & 3.44 & 118 (44)\\
        Nanofabrication & 3.45 & 3.39 & 3.71 & 118 (44)\\
        Dust trap & 3.67 & 3.54 & 3.85 & 47 (30)\\
        IBM Q. Experience & 3.37 & 3.10 & n/a & 57 (0)\\
    \end{tabular}
    \caption{Average participant ratings of the lab activities at USEQIP with more than two iterations from 2017--2025 on a four-point Likert scale from Strongly Agree (4) to Strongly Disagree (1). The questions asked are detailed in Sec.~\ref{sec:feedback}. N corresponds to the number of students who evaluated their interest and the perceived relevance of each lab, with the number in brackets corresponding to the number who evaluated the novelty of the activity.}
    \label{tab:labfeedback}
\end{table}

Students were also asked a series of questions about the program as a whole to evaluate if it was meeting its goals of creating a sense of belonging and encouraging students to continue in QIST. Using the same 4-point Likert scale, students from 2017--2025 were asked about the uniqueness of the program (``The things I learned in USEQIP, I could not have learned anywhere else in my undergraduate studies''), their excitement during the program (``The program was engaging and exciting''), and the practicality of the program (``I'll apply what I learned at USEQIP in my future academic studies''). From 2023--2025, a question was added about the sense of community developed during the program (``USEQIP showed me that I am part of a larger community of like-minded and smart individuals''). Once again, on average the responses were highly positive, showing that students were excited about what they learned and felt it was relevant to their studies and careers moving forward. In an open-ended portion of the post-workshop survey, many students left comments, such as:

\begin{itemize}
    \item ``USEQIP was an absolutely amazing experience. Not only were the lectures and labs engaging, it was the best chance to spend time with 30 other top students from around the world. The curiosity that everyone showed - from us undergrads to the tenured professors - really excites me for seeing a quantum computing revolution in our near future.'' (2018)
    \item ``It pushes you to a new level that you haven't experienced before in your undergraduate. You get thrown into the deep end and get exposed to so many interesting topics that are super complex, but the lecturers and lab techs do a fantastic job of keeping you afloat enough so that you're learning the entire time. It's a no pressure environment, i.e., no tests, so it's all about what you make of it, what you want to learn is what you will learn. It was amazing getting to know other people with similar interests to me, and make new friends who will help me with my research journey.'' (2023)
    \item ``It was great to experience this program right before starting master's degree, as it provided a good overview of different quantum platforms and topic. This is a great way to evaluate which sounds most interesting/promising to you personally and what do you want to focus on during your further studies.'' (2022)

\end{itemize}

\begin{table}
    \centering
    \begin{tabular}{c|c|c|c|c|c}
         & Unique  & Exciting  & Practical & Community & N \\
           & (/4) & (/4) & (/4) & (/4) & \\
        \hline
        2017 & 3.33& 3.44 & 3.39 & n/a & 17 \\
        2018 &  3.52 & 3.83 & 3.70 & n/a & 23\\
        2019 &  3.65 & 3.94 & 3.65 & n/a & 17\\
        2022 &  3.29 & 3.76 & 3.65 & n/a & 17\\
        2023 &  3.10 & 3.71 & 3.48 & 3.67 & 17\\
        2024 &  3.50 & 3.71 & 3.54 & 3.57 & 14\\
        2025 &  2.92 & 3.46 & 3.08 & 3.15 & 13 \\
        \hline 
        Average& 3.33 & 3.69 & 3.50 & 3.46 & 118 (44)\\
    \end{tabular}
    \caption{Average participant ratings of the USEQIP program by year based on four agree/disagree questions detailed in Sec.~\ref{sec:feedback} on a four-point Likert scale. N is the number of survey responses, with the number in brackets being applicable to the ``Community'' question only.}
    \label{tab:programfeedback}
\end{table}

\subsection{Alumni analysis}

\begin{figure}
    \centering
    \includegraphics[width=\columnwidth]{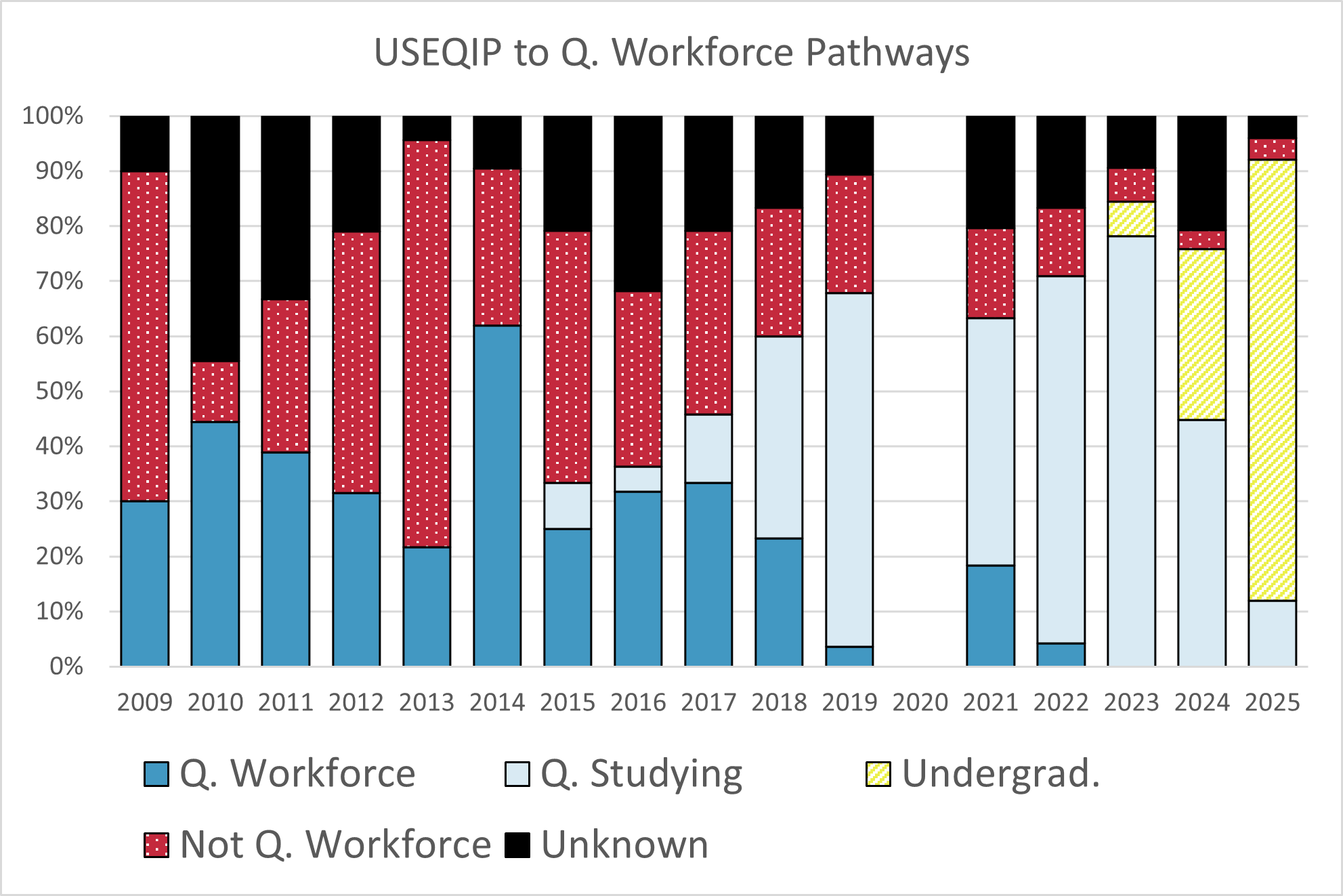}
    \includegraphics[width=\columnwidth]{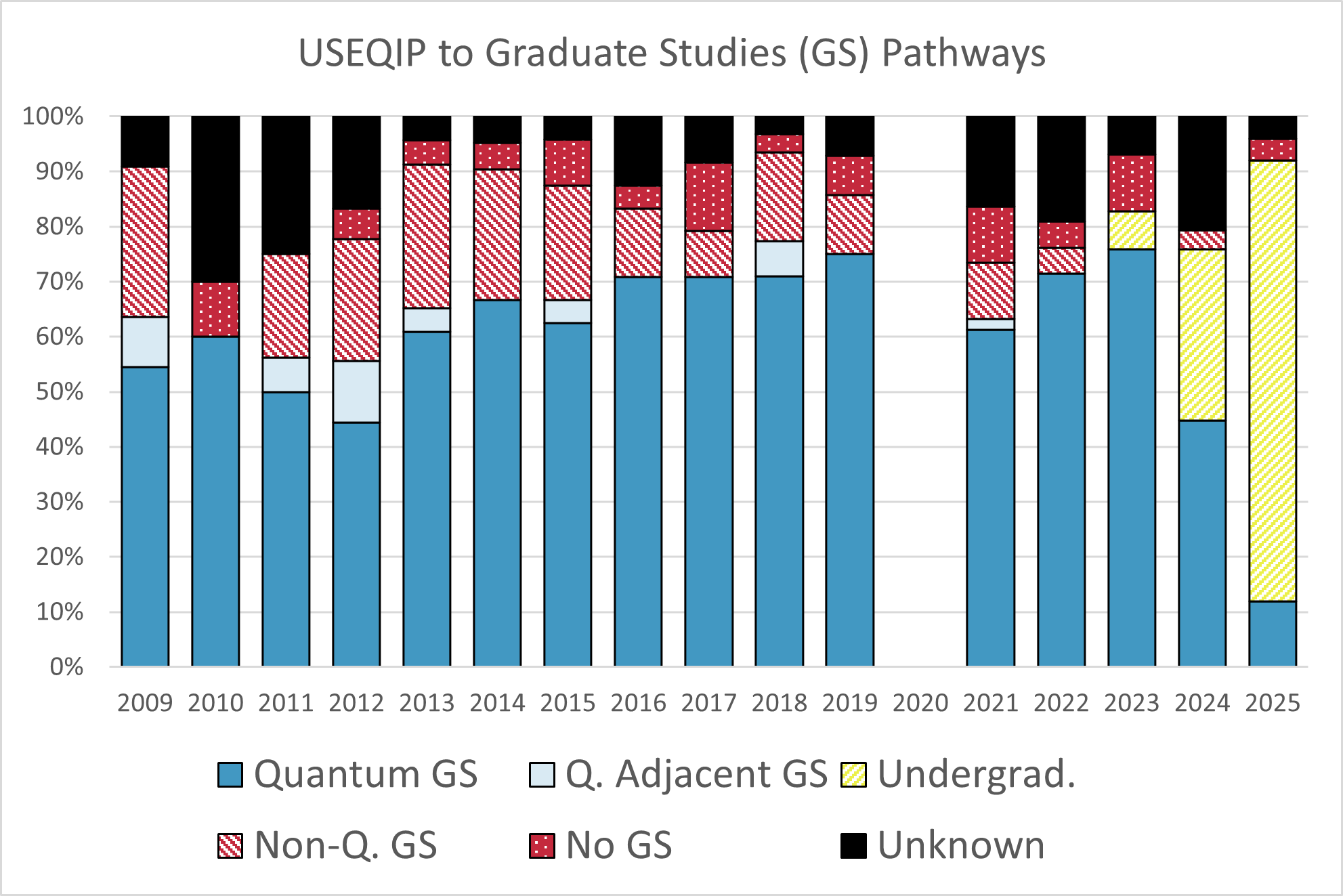}
    \caption{Quantum workforce status (top) and graduate studies (GS) pathways (bottom) of alumni from the USEQIP program grouped by the year they participated, from a digital survey of their public online presence as of January 2026.}
    \label{fig:alumnichart}
\end{figure}

To gauge whether program alumni stay in the QIST field, we classified program alumni from 2009--2025 based on a digital survey of publicly available information on their professional websites and professional social media profiles (e.g., LinkedIn). Information was found for over 82\% of USEQIP alumni (320/387) in January 2026. Of those 312, 77 (20\%) have a role in the quantum workforce and 114 (30\%) are in graduate studies consistent with quantum research activities. Excluding those for whom no information could be found and those who remain in undergraduate studies, 66\% of students continued in quantum-related fields, either in ongoing studies or as their career post-graduation. The year-by-year breakdown is seen in Fig.~\ref{fig:alumnichart}. We also looked at what fraction of the students went into graduate studies researching quantum technology, irrespective of whether they stayed in QIST after graduation. 75\% (231/309) of program alumni who have completed undergraduate degrees and about whom information could be found continued on in graduate studies related to quantum science and technologies, which has held reasonably steady for many years, as seen in Fig.~\ref{fig:alumnichart}. 53 of those students (14\%) returned to the University of Waterloo specifically for their graduate studies. Some topics related to quantum science, such as materials sciences, were labeled as ``quantum-adjacent'' and counted separately.

\subsection{\label{survey}Alumni survey}

In early 2026, a survey was sent to all 198 program alumni from 2018 to 2025, receiving 36 responses. Alumni were asked to describe their field of work or study as ``Quantum information science and technology (QIST),'' ``Quantum-adjacent science and technology,'' ``Science, technology, engineering, or mathematics (STEM) unrelated to quantum,'' or ``non-STEM.'' Of those who had completed their undergraduate studies, 18/33 (55\%) remained in QIST, with 9/33 (27\%) in a field they identified as quantum-adjacent. Alumni were also asked, ``How much do you feel USEQIP influenced your academic and/or career path?'' with options of ``A great deal/a lot,'' ``a moderate amount,'' ``a little,'' and ``not at all.'' In total, 25/36 (69\%) felt USEQIP influenced their career a great deal, with only 4/36 saying a little or not at all. If we remove responses from the non-traditional virtual program alumni of 2021, all students felt it influenced their pathway a moderate amount (5/26) or a great deal (21/26), indicating that the in-person interactions and peer networking are important. Focusing strictly on students who have finished undergraduate studies, remained in QIST, and excluding 2021, 11/12 responded that USEQIP influenced their career a great deal, and 1/12 indicated a moderate amount.

In the same survey, alumni were asked to rank which aspects of the program they felt were most important and impactful in retrospect, choose between ``Lectures,'' ``Hands-on lab activities,'' ``Peer networking,'' and ``Networking with professors and researchers,'' with a ranking of 1 indicating most important and 4 least important. Rankings were generally close, with average rankings of 2.97 for networking with researchers, 2.44 for lectures, 2.44 for peer networking, and 2.15 for hands-on labs. When asked to rate their overall USEQIP experience with the benefit of retrospect on a five-point Likert scale from Strongly Positive to Strongly Negative, 18/36 responded with Strongly Positive, 16/36 with Positive, 2/36 with Neutral (both from the 2021 virtual program), and none as Negative or Strongly Negative. 

Open-ended feedback elicited responses including:

\begin{itemize}
\item ``USEQIP and URA gave me a very important jumpstart towards my dream career of a quantum information scientist. People I have met at IQC during USEQIP/URA and after, as a Master's student, still form a core of my network and friend group in Canada.'' (2022 alum)
\item ``It was very cool to do the experiments and learn how things really work, especially as a theorist and mathematician who usually wouldn't get those experiences.'' (2022 alum)
\item ``USEQIP was a great experience to connect with other people interested in getting into quantum and getting access to highly specialized education. I come from [South America], and at the time I did not know anyone around interested in quantum computing or working in the field. It was a privilege to attend and the experienfe [sic] shaped my career.'' (2018 alum)
\item ``I still talk to friends whom I made at USEQIP; I think that was the most important aspect of it for me. That, and exposure to different topics in quantum computing. It's not that I walked out of USEQIP feeling like an expert in anything I learned about, but I walked out knowing that certain quantum computing systems existed and that was useful for my early understanding of what I could do research in.'' (2019 alum)
\item ``USEQIP was one of the most valuable experiences I had in my undergraduate degree. Being able to spend time with peers from across the world in Waterloo gave me a lot of inspiration and insight into what I wanted to do with my future! I ended up pursuing a quantum topic for my Master's degree as a result, which is likely not something I would have considered without USEQIP.'' (2018 alum)
\end{itemize}

Some criticism of the program from the open-ended feedback included limited direct interaction with professors and a desire for more organized interactions throughout the summer for those who stayed for a research internship.

\section{Conclusion}
The rapidly increasing demand for a quantum-literate and quantum-proficient workforce has demanded a steady rise in educational activities and opportunities available for students at all levels. The USEQIP model provides students with hands-on learning experiences with quantum equipment capable of showing quantum advantage, internship opportunities with leading researchers, and a peer network to foster a sense of belonging. As detailed in Sec.~\ref{sec:students}, an open-tent approach that provides opportunities for students from varied disciplinary backgrounds and international locations, as well as structural considerations to include under-represented groups, builds a strong social network for students and helps strengthen a more diverse talent pipeline. Dedicated, purpose-built space and equipment allowed for the development of in-depth and interactive activities, as detailed in Sec.~\ref{sec:labs}. USEQIP alumni are highly likely to continue graduate studies in QIST and join the quantum workforce afterwards, as seen in Sec.~\ref{sec:feedback}. As QIST evolves, so too must programs like USEQIP. By focusing on a core set of experimental tools and skills, paired with thoughtful lectures and workshops from researchers and industry partners, USEQIP will continue to act as an enabling pathway for undergraduate students of varied backgrounds to enter quantum research and the quantum workforce.


\section*{Acknowledgments}

We are grateful to support from the Canada First Research Excellence Fund (CFREF) under the Transformative Quantum Technologies (TQT) program, Innovation Science and Economic Development (ISED) Canada under the Strategic Science Fund (SSF), MITACS, the Canada Excellence Research Chairs (CERC) program, and Mike and Ophelia Lazaridis. We thank R.~Islam, L.~Hahn, and the Quantum Information with Trapped Ions group for the dust trap lab mentioned in Sec.~\ref{sec:labs:other}, and C.~T.~Earnest for the chip design used in the lab detailed in Sec.~\ref{sec:labs:nanofab}. We are immensely grateful to the numerous faculty and graduate students at the Institute for Quantum Computing who have supported USEQIP by leading labs, delivering lectures, and supervising student projects, and the staff who have supported the administration and management of the program, including K.~Kuntz and many others.


\bibliography{useqip}

\end{document}